\documentclass[aps,prb,amsmath,amssymb,superscriptaddress,reprint, a4paper] {revtex4-1}
\usepackage{graphicx}
\usepackage{hyperref,amsmath}
\usepackage{amsmath}
\usepackage{dcolumn}
\usepackage{bm}
\usepackage{newfloat}
\usepackage{multibib}
\usepackage{color}

\begin{document}
\title{Injection locking of an electro-optomechanical device}

\author{Christiaan Bekker}
\affiliation{Centre for Engineered Quantum Systems, School of Mathematics and Physics, The University of Queensland, Australia}
\author{Rachpon Kalra}
\affiliation{Centre for Engineered Quantum Systems, School of Mathematics and Physics, The University of Queensland, Australia}
\author{Christopher Baker}
\affiliation{Centre for Engineered Quantum Systems, School of Mathematics and Physics, The University of Queensland, Australia}
\author{Warwick P. Bowen}
\affiliation{Centre for Engineered Quantum Systems, School of Mathematics and Physics, The University of Queensland, Australia}


\maketitle

Advances in optomechanics have enabled significant achievements in precision sensing and control of matter, including the detection of gravitational waves and the cooling of mechanical systems to their quantum ground state. Recently, the inherent non-linearity in the optomechanical interaction has been harnessed to explore synchronization effects, including the spontaneous locking of an oscillator to a reference injection signal delivered via the optical field. Here, we present the first demonstration of a radiation-pressure driven optomechanical system locking to an inertial drive, with actuation provided by an integrated electrical interface. We use the injection signal to suppress drift in the optomechanical oscillation frequency, strongly reducing phase noise by over 55~dBc/Hz at 2~Hz offset. We further employ the injection tone to tune the oscillation frequency by more than 2 million times its narrowed linewidth. In addition, we uncover previously unreported synchronization dynamics, enabled by the independence of the inertial drive from the optical drive field. Finally, we show that our approach may enable control of the optomechanical gain competition between different mechanical modes of a single resonator. The electrical interface allows enhanced scalability for future applications involving arrays of injection-locked precision sensors.




\section{Introduction}

The ability to engineer strong interactions between high-quality electromagnetic cavities and mechanical resonators has led to a rich variety of results in the field of optomechanics, including ground-state cooling~\cite{Chan2011,Teufel2011}, quantum-limited measurement~\cite{Bagci2014,Arcizet2006a} and optomechanical entanglement~\cite{Palomaki2013b}. Optomechanical systems have further proven to be a powerful tool for applications involving precision sensing~\cite{Aspelmeyer2014,  Hu2013, bowen2015book}. Minute perturbations in the physical environment can be detected through changes in the resonance frequency of a nanomechanical resonator, making these systems widely used as mass~\cite{Jensen2008,Chaste2012} and gas~\cite{Bargatin2012} sensors. Furthermore, many optomechanical systems can be fabricated on-chip, allowing for dense arrays of sensors to be employed for applications such as multi-particle sensing for the detection of lung cancer~\cite{Fernandes2015}. One important feature of optomechanical systems is the ability of the optical field to amplify mechanical motion. This regenerative amplification has been shown to allow enhanced sensitivity~\cite{feng2008nnano} and presents an avenue for the exploration of synchronization phenomena in optomechanical systems. 

Non-linear effects can cause two resonators to become synchronized, where the phase of their individual oscillations lock with respect to each other. A network of coupled oscillators can spontaneously synchronize, as was first reported for pendulum clocks hung from a common frame~\cite{Bennett2002} and further observed in numerous biological systems including the flashing of fireflies and the chirping of crickets~\cite{mirollo1990siam}. Alternatively, a single oscillator can synchronize to the phase of an externally applied drive in an effect known as injection locking, as has been studied extensively in the context of electrical tank circuits~\cite{adler1946,Paciorek1965,Razavi2004}, implemented in non-linear mechanical resonators \cite{Seitner2017} and further observed for the effect of light on human circadian rhythms~\cite{Duffy2009}. Both these forms of synchronization have been explored in optomechanical systems. Arrays of optomechanical systems can synchronize when coupled through overlap of their optical modes~\cite{Zhang2012,Zhang2015a}, via coupling to a common optical waveguide~\cite{Gil-Santos2017} or through direct inertial coupling~\cite{Heinrich2011,Ludwig2013}. Injection locking of a single optomechanical oscillator has also been demonstrated~\cite{Hossein-Zadeh2008,Shah2015,Shlomi2015,Zalalutdinov2003}.

Conventionally, the injection signal is delivered to the optomechanical oscillator through the optical field, achieved experimentally via modulation of the input laser power~\cite{Hossein-Zadeh2008,Shah2015,Shlomi2015}. Alternatively, as pointed out in a recent theoretical study~\cite{amitai2017pra}, the oscillator can be driven with a direct inertial force. By evading the cavity filtering inherent to optical driving, this introduces different synchronization dynamics. Moreover, inertial drives can be locally applied to individual oscillators with integrated electrodes, thus presenting a scalable approach for applications utilizing arrays of oscillators. While direct inertial forcing has been achieved in a bolometric system~\cite{Zalalutdinov2003}, it has not previously been demonstrated with a radiation-pressure optomechanical oscillator. The conservative nature of the radiation-pressure interaction, in contrast to bolometric systems, enables many of the most important applications of optomechanical systems in quantum science and technology~\cite{bowen2015book}. 

In this work, we present the first demonstration of locking of a radiation-pressure optomechanical oscillator to an inertial drive. We demonstrate the ability of the injection signal to suppress phase noise by over 55~dBc/Hz and to tune the oscillation frequency by more than 2 million times the oscillation linewidth, and explore previously unreported locking dynamics. With our approach, the inertial injection signal is the electrostatic force between two electrodes directly patterned onto the body of the resonator. The feed-forward stabilization achieved with this electrical injection locking may prove more scalable for arrays of sensors or complex many-body quantum networks as compared to optical injection locking, or to implementing feedback circuitry for individual free-running resonators~\cite{lee2010prl}. For single oscillator sensors, a potential application of the inertial drive is to employ anomalous cooling~\cite{Kemiktarak2014} to suppress undesired regenerative oscillation of certain mechanical modes, an issue encountered in precision optical sensors such as gravitational wave detectors~\cite{Braginskii2002,Abdi2011,Vyatchanin2012}.

\begin{figure}[t!]
	\includegraphics[width = \linewidth]{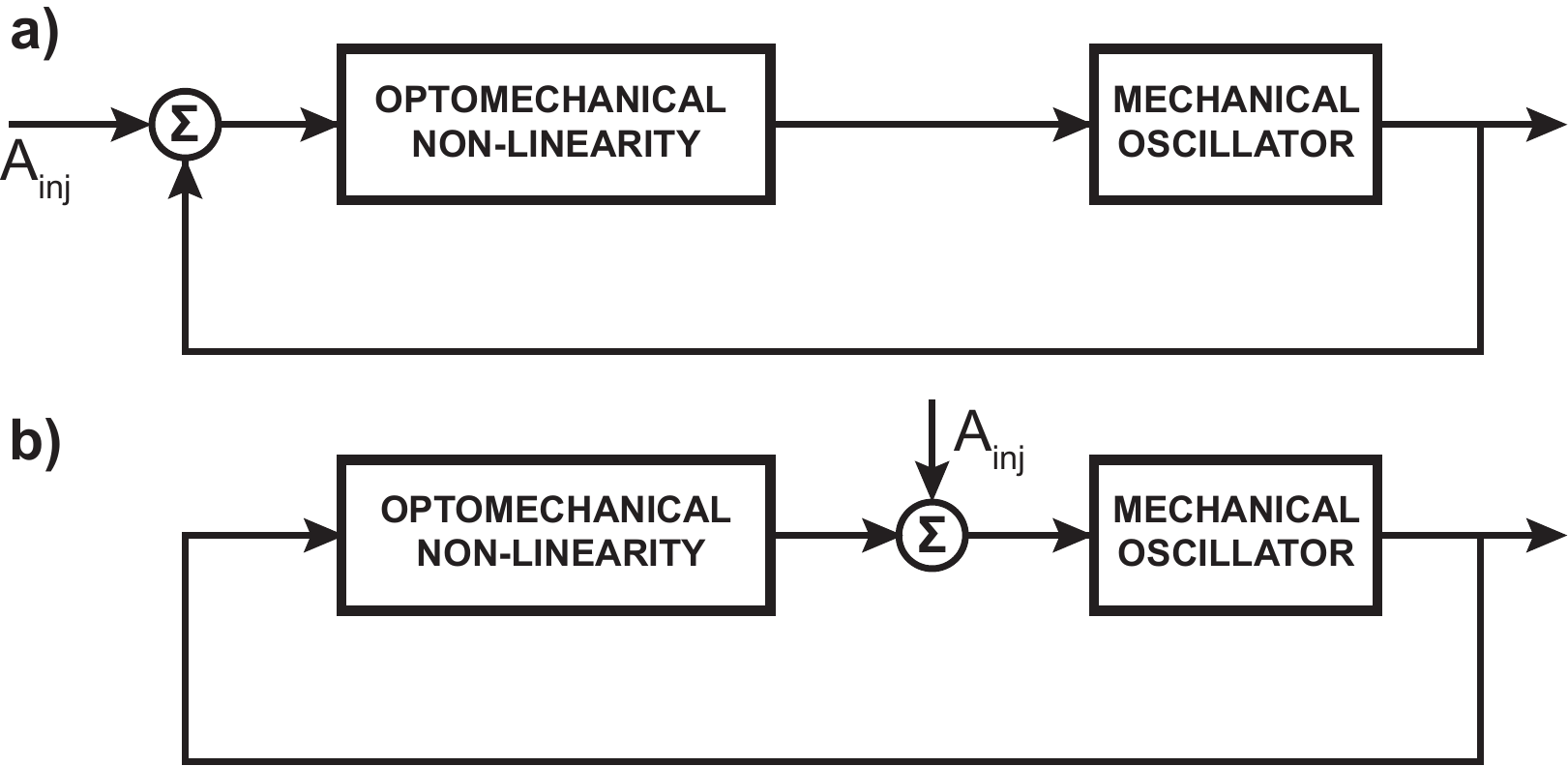}
	\caption{\label{fig:Block} Block diagrams of optomechanical systems where an injection signal $\mathrm{A_{inj}}$ is delivered \textbf{(a)} via the optical mode or \textbf{(b)} as an inertial drive directly to the mechanical resonator. Note that the optical pumping of the cavity is omitted in the schematic.}
\end{figure}

Regenerative oscillation, often referred to as self-sustained oscillation, is maintained via an intrinsic feedback loop introduced by the radiation-pressure interaction, which induces dynamical backaction between the light in the optical cavity and the motion of the mechanical element (see Fig.~\ref{fig:Block}). The synchronization phenomenon studied in this work is enabled by the inherent non-linearity in the optomechanical interaction. As discussed above, all previous injection-locked radiation-pressure optomechanical oscillators have used optical injection, which is described by the same feedback loop (Fig.~\ref{fig:Block}(a)) as widely studied injection-locked tank circuits~\cite{Paciorek1965}. In contrast, inertial injection bypasses the system non-linearity to drive the mechanical oscillator directly, qualitatively modifying the dynamical behaviour of the system. This provides the opportunity to explore different synchronization behaviour. Furthermore, in contrast to optical injection, the inertial force can be made larger than the radiation-pressure force. Our measurements uncover two different regimes of injection locking which display qualitatively different behaviour when the resonator is partially locked. This may stem from the unique way in which our injection signal is applied. 

\section{Electro-optomechanical system}

\begin{figure} [t!]
	\includegraphics[width=\linewidth]{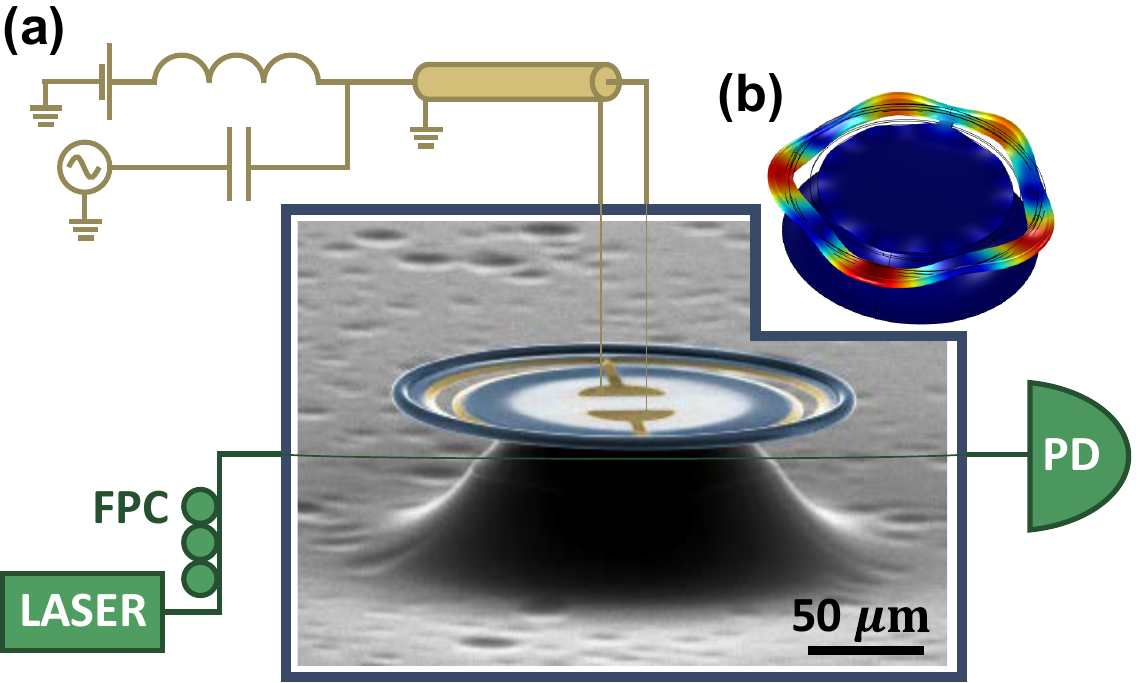}%
	\caption{\label{fig:Setup} \textbf{(a)} Scanning electron micrograph of the silica (blue) microtoroid optomechanical cavity. A circular slot is etched through the device to increase the compliance for radial motion~\cite{Baker2016}. Circular capacitor electrodes (yellow) are patterned on either side of the slot. Components of the optical and electrical measurement setup are shown in green and yellow, respectively. $V_\mathrm{DC}$ and $V_\mathrm{AC}$ are applied to the capacitor via a bias-tee and probe tips controlled by micro-manipulators~\cite{Baker2016}. Laser light is coupled into one end of an optical fiber which feeds into a fiber polarization controller (FPC). The tapered section of the fiber is coupled to the silica WGM using a micro-positioning stage; transmitted light is then collected on a high-speed photodetector (PD). \textbf{(b)} Result of COMSOL simulation showing the `radial breathing mode'-like excitation of such a microtoroid supported by a single spoke.}
\end{figure}

The electro-optomechanical system used in this work consists of a microtoroidal optomechanical oscillator~\cite{carmon2005prl,Taylor2011}  with an integrated electrical interface that allows a radial force to be applied directly to the mechanical resonator. In a previous work, this force was used to demonstrate high-bandwidth tuning of the optical resonance frequency~\cite{Baker2016}. The optical mode is a whispering gallery mode (WGM) of the silica microtoroid of radius 100~$\mu$m, described by its ladder operators $a$ and $a^\dagger$, with resonance frequency $\omega_\mathrm{c}/2\pi \approx 194$~THz and linewidth $\kappa/2\pi\approx100$~MHz. Fig.~\ref{fig:Setup}(a) shows a false-color electron micrograph of the microtoroid device, consisting of a reflown silica disk (blue) atop an etched silicon pedestal (grey). Apart from having a larger radius, this device is identical to that studied in Ref.~\cite{Baker2016}, which can be referred to for details on device fabrication. Measurements are performed by bringing the tapered section of an optical fiber in contact with the microtoroid to couple to the WGM  while reducing taper drift. Laser light of frequency $\omega_\mathrm{L}$ (wavelength~$\sim 1550$~nm) and power $P_\mathrm{in}$ is delivered to one end of the fiber while the transmitted power through the taper $P_\mathrm{out}$ is measured by a photo-detector, allowing the optical WGM to be probed. The optical measurement setup is shown in Fig.~\ref{fig:Setup}(a) in green. All measurements are made in ambient conditions.

The optical mode couples to a mechanical radial breathing mode of the microtoroid, described by its ladder operators $b$ and $b^\dagger$. Fig.~\ref{fig:Setup}(b) shows the result of a finite element method (COMSOL) simulation of the mechanical mode with frequency $\omega_\mathrm{m}/2\pi \approx 8.9$~MHz, effective mass $m_\mathrm{eff}\approx30$~ng and zero-point motion $x_\mathrm{zp} \approx 0.3$~fm. With $\omega_\mathrm{m} / \kappa \approx 0.09$, the optomechanical system operates in the unresolved sideband regime. The strength of the optomechanical coupling is parametrized by the single photon coupling rate $g_0 = G x_\mathrm{zp}$. $G/2\pi\approx 0.9$~GHz/nm, estimated by COMSOL simulations, is the amount by which the radial mechanical motion shifts the optical resonance frequency~\cite{Aspelmeyer2014}. Detuning the laser frequency appropriately allows probing of the mechanical motion through its modulation of the transmitted optical power. The Hamiltonian in the frame rotating with the frequency of the laser drive $\omega_\mathrm{L}$ is

\begin{equation}
H_\mathrm{om} = \hbar \omega_\mathrm{m} b^\dagger b - \hbar \Delta a^\dagger a - \hbar g_0 a^\dagger a (b+b^\dagger) +  \hbar A_\mathrm{L} (a+a^\dagger).
\end{equation}

\noindent Here, $\Delta = \omega_\mathrm{L} - \omega_\mathrm{c}$ is the detuning and $A_\mathrm{L} = \sqrt{\eta P_\mathrm{in}\kappa_\mathrm{ex}/\hbar\omega_\mathrm{L}}$ is the optical drive amplitude, where $\kappa_\mathrm{ex}$ is the optical coupling rate between the taper and the WGM and $\eta\approx0.15$ captures the power lost from the laser to the taper. In addition to allowing measurement of the mechanical motion, when blue-detuned, the laser drive applies a radiation-pressure force that amplifies the mechanical motion. There exists a threshold for $P_\mathrm{in}$ beyond which the amplification of the mechanical motion provided by the light exceeds the intrinsic loss rate of the mechanical mode, resulting in regenerative mechanical oscillations~\cite{kippenberg2005prl, Aspelmeyer2014}. 

\begin{figure} [!t]
	\includegraphics[width=\linewidth]{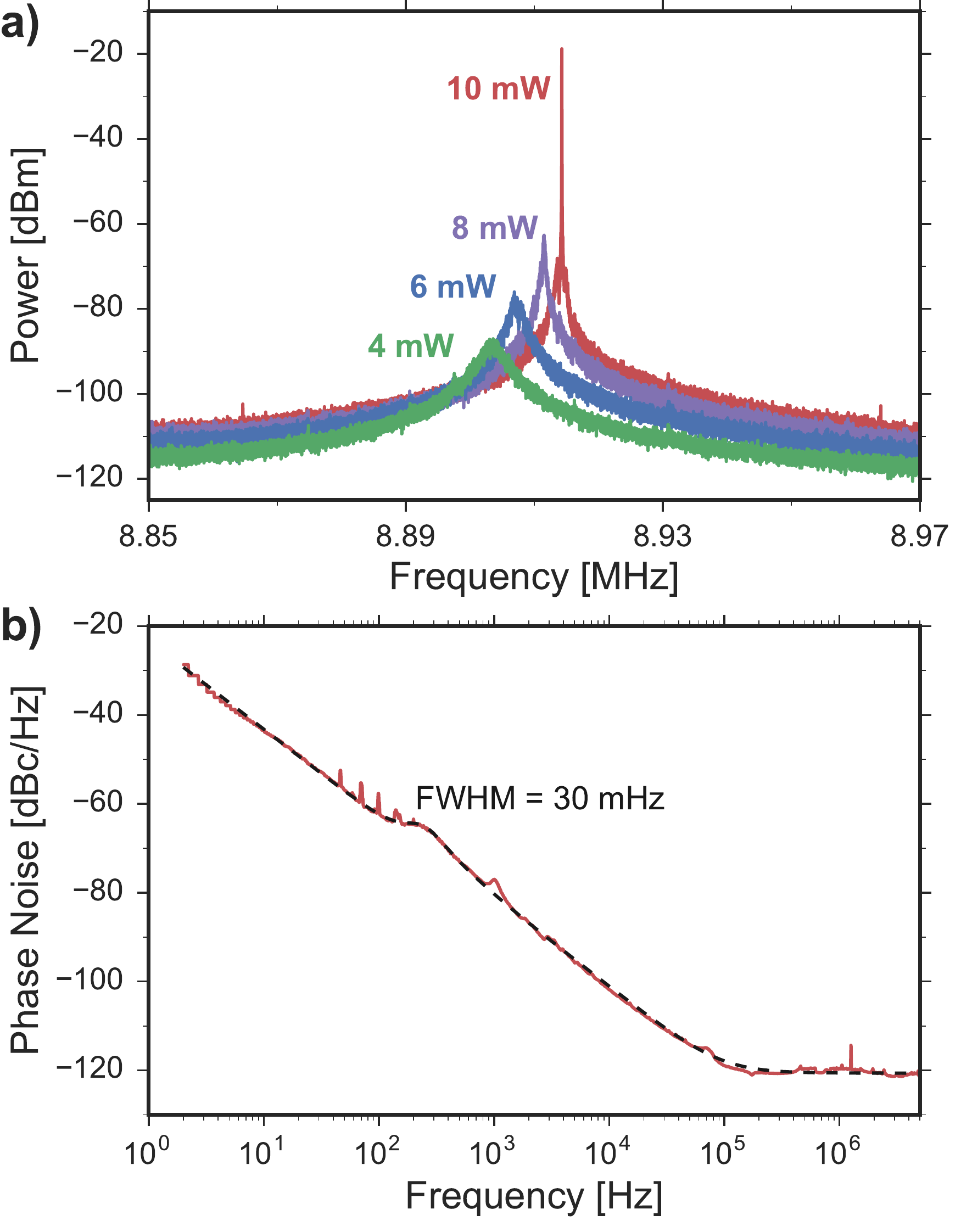}%
	\caption{\label{fig:Regen} \textbf{(a)} Power spectra of the mechanical mode measured via its modulation of the transmitted optical power $P_\mathrm{out}$ for $P_\mathrm{in}=4$, 6, 8 and 10~mW. For each setting of $P_\mathrm{in}$, the laser detuning (in the range of $\kappa/2$) is modified to maximize the mechanical modulation of $P_\mathrm{out}$. At low input powers $P_\mathrm{in}$, the mechanical motion is dominated by thermal excitation. At 10~mW, regenerative oscillation is observed, marked by a significant increase in oscillation amplitude and linewidth narrowing. \textbf{(b)} Phase noise measurement of the mechanical oscillations with $P_\mathrm{in}=10$~mW. A fit to the data shows that the linewidth of the regenerative oscillations is 30~mHz. Note that we add to our fit a Lorentzian peak at an offset of 220~Hz and the noise floor at $-120$~dBc/Hz.}
\end{figure}

Fig.~\ref{fig:Regen}(a) shows the measured power spectrum of the radial breathing mode for varying input power $P_\mathrm{in}$. Operating in the unresolved sideband regime, the laser frequency $\omega_\mathrm{L}$ is tuned at each power setting to maximize the mechanical modulation of $P_\mathrm{out}$. For $P_\mathrm{in} = 4$~mW, thermal excitation of the mode dominates its motion and the intrinsic mechanical linewidth (full-width at half-maximum, or FWHM) $\Gamma/2\pi\approx 15$~kHz can be measured. As $P_\mathrm{in}$ increases, the mechanical mode narrows and is frequency-upshifted, consistent with optomechanical theory~\cite{Aspelmeyer2014, bowen2015book}. Once $P_\mathrm{in}$ surpasses the threshold for regenerative oscillation, the oscillation amplitude increases by several orders of magnitude and the linewidth narrows beyond the frequency resolution of the spectrum analyzer, as shown for $P_\mathrm{in}=10$~mW. At this point, the amplitude is limited by the inherent non-linearity in the optomechanical interaction that drives the oscillations and enables the synchronization effects studied in this work. This non-linearity can be intuitively understood by considering the Lorentzian lineshape of the optical mode as a function of the detuning caused by mechanical displacement~\cite{Aspelmeyer2014, bowen2015book}. For small oscillation amplitudes, the power coupled into the cavity varies linearly with mechanical displacement. However, at larger amplitudes comparable or greater to $\kappa / G$ that probe the Lorentzian shape of the mode, this is no longer the case. Any mechanical non-linearity can be neglected given the relatively small oscillation amplitudes studied here, estimated to be $< 100$~pm, several orders of magnitude smaller than the smallest feature in the device. Kerr and Raman optical non-linearities can also be neglected as the incident powers used are orders of magnitude below the threshold required to observe these effects~\cite{kippenberg2004prl}.

The electrical interface of the device allows for an inertial force to be applied to the microtoroid that is completely independent of the optics~\cite{Baker2016, lee2010prl}. This is achieved through the integration of a circular capacitor (highlighted in yellow in Fig.~\ref{fig:Setup}(a)) with capacitance $C\approx7$~fF, whose attractive force occurs in the radial direction and hence has strong overlap with the radial breathing mode. As shown in the figure, a circular slot is etched through the disk, between the gold electrodes, to increase mechanical compliance. A voltage bias is applied across the capacitor via tungsten probe tips. Ref.~\cite{Baker2016} studied the ability of an applied voltage to tune the optical resonance frequency and did not operate in a regime where optical amplification of the mechanical motion was appreciable. By contrast, here we seek to enhance the optomechanical gain by increasing the laser power inside the cavity and operating with a blue-detuned laser. This allows us to reach the regime of regenerative oscillation and, by applying an RF voltage with frequency $\omega_\mathrm{d} \approx \omega_\mathrm{m}$, study synchronization of the mechanical oscillations. 

We apply a drive voltage $V_\mathrm{d}(t)= V_\mathrm{DC} + V_\mathrm{AC}\cos(\omega_\mathrm{d} t)$. The DC and $2\omega_\mathrm{d}$ components of the resulting attractive force between the capacitor plates (proportional to $V_\mathrm{d}^2$~\cite{Baker2016}) are off-resonant and can be omitted to obtain the drive force $F_\mathrm{d}(t)=(\delta C / \delta x)V_\mathrm{DC}V_\mathrm{AC}\cos(\omega_\mathrm{d} t)$. The change in capacitance per unit mechanical displacement in our device is estimated to be $\delta C / \delta x\approx4\times10^{-4}$~fF/nm from finite element method simulations. The complete system Hamiltonian is then $H=H_\mathrm{om}+H_\mathrm{d}$, where

\begin{equation}
H_\mathrm{d}=x_\mathrm{zp}F_\mathrm{d}(t)(b+b^\dagger).
\label{eq:Drive}
\end{equation}

\noindent As discussed in the previous section, we highlight that this form of drive is distinct from previous demonstrations of locking in optomechanical systems which use optical radiation-pressure modulation driving~\cite{Hossein-Zadeh2008, Shlomi2015}. While the two implementations require similar instrumentation overhead for the synchronization of a single oscillator, our electrical approach should be more scalable for applications involving arrays of oscillators. A single electrical drive can be straightforwardly distributed to all resonators on a chip, whereas the optical driving scheme would generally require one optical modulator per resonator due to mismatches in optical resonance frequencies. The electrical approach also benefits from \textit{in situ} tuning of the optical resonance frequency with a DC voltage~\cite{Baker2016}.

\section{Injection locking}

\subsection{Locking and Stability}

Fig.~\ref{fig:Regen}(a) shows the power spectrum of the regeneratively oscillating mechanical mode. We see that while the natural linewidth of the mechanical mode is 15~kHz, the optomechanical gain significantly narrows the linewidth such that it is no longer resolved by the spectrum analyzer. We thus use a phase-noise analyzer to capture the narrowed linewidth. Fig.~\ref{fig:Regen}(b) shows the measured phase noise for the optomechanical oscillator with $P_\mathrm{in}=10$~mW. The expected phase noise (units of dBc/Hz) for the oscillator is~\cite{demir2000ieee}

\begin{equation}
\mathcal{L}(f) = 10\log_{10} \left( \frac{1}{\pi}\frac{f_\mathrm{hw}}{(f_\mathrm{hw})^2+(\Delta f)^2} \right),
\end{equation}

\noindent where $\Delta f$ is the frequency offset from the carrier and $f_\mathrm{hw}$ the half-width at half-maximum. The data in Fig.~\ref{fig:Regen}(b) is fit to the above equation, confirming the expected $1/(\Delta f)^2$ lineshape and yielding a linewidth of 30~mHz. This value is $5\times10^5$ times smaller than the intrinsic linewidth of the mechanical mode.

\begin{figure} [!t]
	\includegraphics[width=\linewidth]{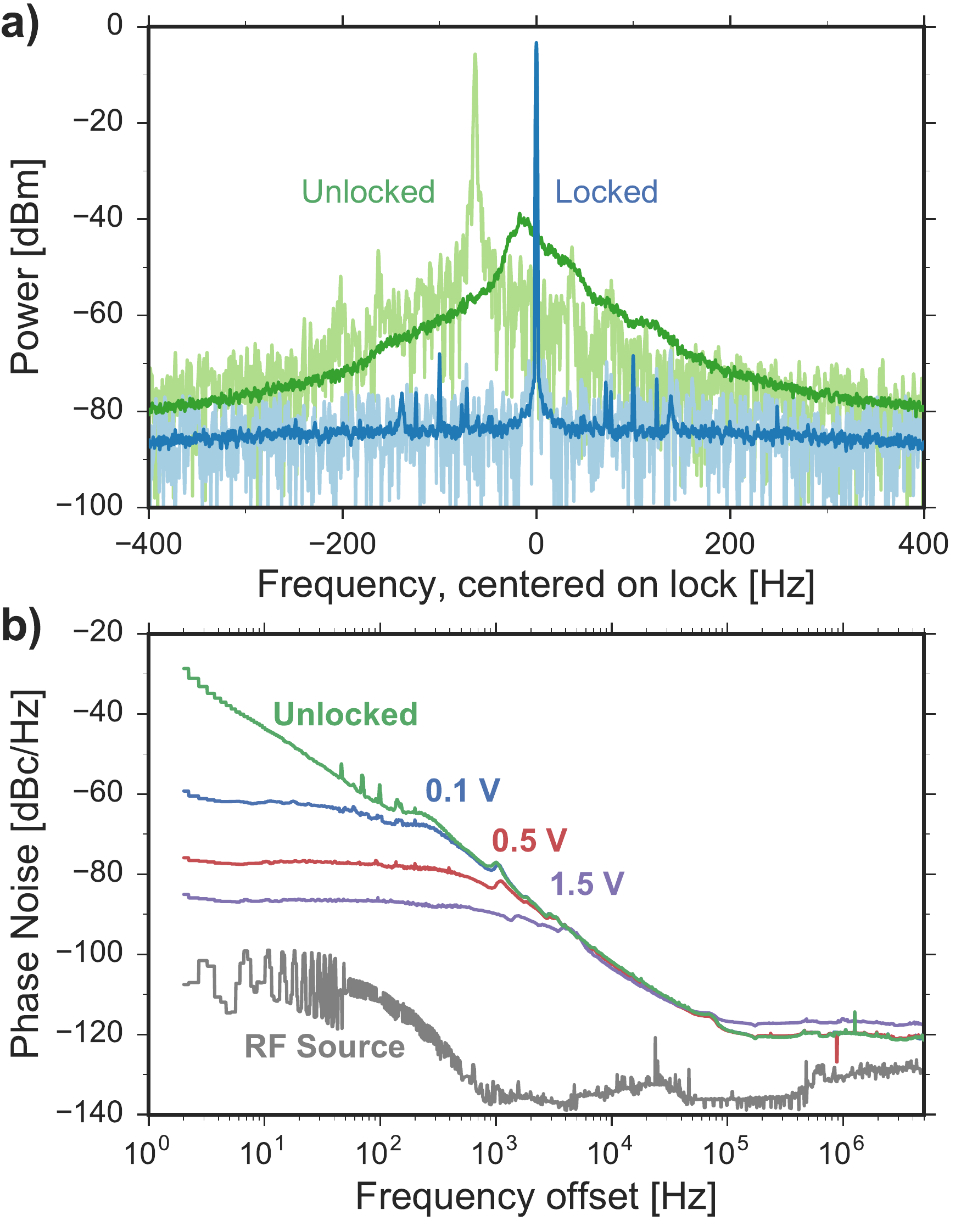}%
	\caption{\label{fig:Locking} \textbf{(a)}  Comparison of the power spectra of unlocked (green) and locked (blue) regenerative oscillations with $P_\mathrm{in}=20$~mW. The light and dark traces correspond to single acquisitions and averages of 50 consecutive acquisitions, respectively, presented to highlight the effect of drift in the mechanical oscillation frequency. \textbf{(b)} Comparison of phase noise traces for varying $V_\mathrm{AC}$ with $P_\mathrm{in}=10$~mW and the direct phase noise of the RF signal generator. Higher drive strengths result in a greater suppression of mechanical phase noise over a wider range of frequency offsets.}
\end{figure}

While the linewidth of the regenerative oscillations is extremely small, this particular measurement was performed with frequency offsets above 2~Hz and therefore does not capture the fact that the oscillation frequency drifts significantly in the timescale of minutes. This drift would be detrimental to the use of the oscillator for applications that require long-term frequency stability such as mass or temperature sensors. To highlight the broadening caused by this drift, we compare a single acquired power spectrum (light green) to the average of 50 consecutively acquired traces acquired over a timescale of minutes (dark green) for $P_\mathrm{in}=20$~mW in Fig.~\ref{fig:Locking}(a). The oscillation frequency drifts over a range of $\sim 100$~Hz in this time, corresponding to more than 3000 times the reduced linewidth. This significant drift can be eliminated by applying the external drive described by Eq.~\ref{eq:Drive} to which the mechanical oscillations can synchronize.

To demonstrate this effect, we apply a drive $V_\mathrm{AC}=0.5$~V with $V_\mathrm{DC}=50$~V at a frequency $\omega_\mathrm{d}/2\pi=8.914$~MHz. Fig.~\ref{fig:Locking}(a) shows a single trace (light blue) and an average of 50 acquired traces (dark blue) of the power spectrum of the locked oscillations. We observe that the peak in the averaged trace overlaps exactly with that of the single acquisition, indicating that the oscillations are indeed locked and that the frequency drift is eliminated. In addition to the elimination of frequency drift, a comparison of the single acquisitions of the locked (light blue) and unlocked (light green) traces reveals that the locking significantly reduces the quasi-instantaneous linewidth. This effect of injection locking is well known in other types of synchronized oscillators~\cite{Razavi2004}, and has been demonstrated before in optomechanics with a radiation-pressure-modulated drive~\cite{Hossein-Zadeh2008}. It is also used, for instance, in laser systems to reduce the phase noise of a noisy high-power laser by locking it to a phase-stable low-power laser~\cite{buczek1973ieee}. 

To explore this further, we measure the phase noise for the locked oscillations for varying drive strengths. Fig.~\ref{fig:Locking}(b) plots the phase noise for $V_\mathrm{AC}=0.1$~V, 0.5~V and 1.5~V with $P_\mathrm{in}=10$~mW. Also included in the figure is the phase noise trace of the unlocked oscillations for comparison (reproduced from Fig.~\ref{fig:Regen}(b)) as well as the phase noise of the RF voltage source itself. We clearly observe a significant suppression of phase noise when the oscillator is locked, with a maximum suppression of over 55~dBc/Hz at 2~Hz. A stronger drive results in greater phase noise suppression, approaching the noise of the RF source itself~\cite{Hossein-Zadeh2008}. As discussed in Ref.~\cite{Razavi2004} in the context of electrical tank circuits, and shown in Fig.~\ref{fig:Locking}(b), the phase noise is suppressed up to a critical frequency offset, which depends on the drive strength. Above that frequency, the phase noise of the unlocked or free-running oscillator is recovered. As we will see in the next section, where injection locking is used to tune the frequency of regenerative oscillations, this critical frequency corresponds to the locking range for the particular drive strength.

\subsection{End of Lock Range Dynamics} 

As well as stabilising the mechanical oscillations, the injection signal can also be used to tune the oscillation frequency. The range of frequencies over which the mechanical mode can be locked has been studied extensively in the context of electrical tank circuits~\cite{adler1946, Razavi2004} and, more recently, optomechanical systems~\cite{amitai2017pra}. By definition, the oscillator is locked to the drive when the phase difference between it and the external drive, $\Delta\phi$, is stationary in time. When locked, $\Delta\phi$ is zero at the natural mechanical resonance frequency, and grows with detuning to $\pm\pi/2$ at the end of the lock range, after which point the oscillator fails to lock~\cite{adler1946}. This change in phase with mechanical resonance frequency allows an injection-locked oscillator to be used for sensing applications where shifts in the mechanical resonance frequency are to be detected. This may offer benefits for sensor arrays compared to other approaches, such as using free-running regenerative oscillators or phase-locking of driven oscillators, as stability is achieved without the need for individual feedback control circuitry to each sensor.

Here, we explore the dependence of lock range on the two independent variables in our system: (i) the drive strength and (ii) the optical power, and show good agreement with the recent theoretical study performed by Amitai~\textit{et al.}~\cite{amitai2017pra}. In addition to this, we find that previously studied end-of-lock-range dynamics only occur for sufficiently high optical power, with a new class of behaviour evident at lower optical powers.

\begin{figure} [!t]
	\includegraphics[width=\linewidth]{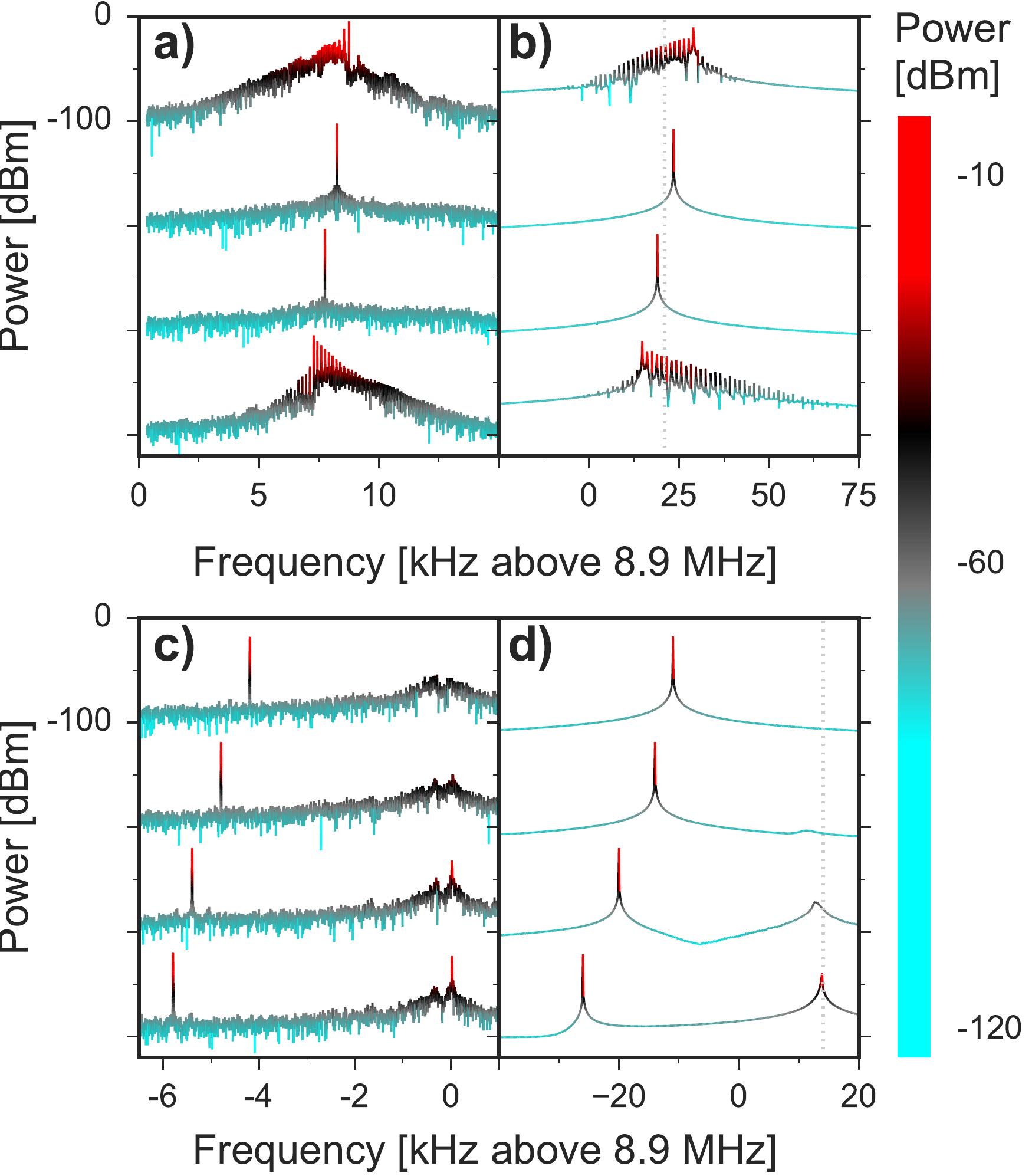}
	\caption{\label{fig:Edgeoflock} Comparison of the end of lock range dynamics for the two observed regimes. \textbf{(a)} Mechanical power spectra of regenerative oscillations at $P_\mathrm{in}=23$~mW with $V_\mathrm{DC}=50$~V and a $V_\mathrm{AC}=0.5$~V injection signal for varying $\omega_\mathrm{d}$. Each consecutive trace corresponds to a shift of $-500$~Hz in $\omega_\mathrm{d}$ and is offset in the graph by -100 dBm for clarity. The edges of the lock range (top and bottom traces) are clearly marked by the characteristic quasi-lock spectrum. \textbf{(b)} Result of numerical solutions of the equations of motion for the electro-optomechanical system with parameters approximating those used in the measurements for sub-figure (a). Good qualitative agreement between experiment and theory of the end-of-lock-range dynamics in the quasi-locking regime is shown. \textbf{(c)} Measured power spectra where $P_\mathrm{in}$ is reduced to 10~mW to achieve a larger lock range. Each consecutive trace corresponds to a shift of  $\approx-500$~Hz in $\omega_\mathrm{d}$, spanning over the lower-end of the lock range which is here marked by the emergence of the peak at the natural resonance frequency. \textbf{(d)} Result of numerical solutions with parameters approximating those used in measurements for sub-figure (c).}
\end{figure}

We begin by setting the system to regeneratively oscillate with $P_\mathrm{in}=23$~mW and applying an AC voltage $V_\mathrm{AC}=0.5~$V at a frequency close to the mechanical resonance. With these settings, the amplitude of the mechanical oscillations due to the radiation-pressure force dominates over the inertial drive by at least an order of magnitude. The oscillations successfully lock to the drive and $\omega_\mathrm{d}$ is then increased (decreased) to find the upper (lower) end of the lock range. The end of the lock range is unambiguously signalled on the spectrum analyzer, marked by a characteristic spectrum that is representative of quasi-locking \cite{Razavi2004}. As discussed in Ref.~\cite{Razavi2004}, in this regime of quasi-locking the oscillator slips in and out of lock, which manifests as $\Delta\phi$ alternating between periods of being held near $\pm \pi/2$ and cycling through $2\pi$ radians before being locked near $\pm \pi/2$ again (c.f. Supplementary Materials). The fraction of the time for which $\Delta\phi$ is held near $\pm \pi/2$ decreases as the drive frequency moves further out of the locking range~\cite{Razavi2004}, until the oscillator is no longer locked at all. The Supplementary Materials cover the phase dynamics in greater detail. Here, it suffices to consider that this results in a phase difference that oscillates in time. This translates to a periodic frequency modulation that results in the side-bands that make up the triangular singled-sided spectrum characteristic of the quasi-locking phenomenon~\cite{Razavi2004}, as shown in Fig.~\ref{fig:Edgeoflock}(a). The top (bottom) trace corresponds to the lower (upper) end of the lock range while the two spectra in between are within the lock range. 

For injection frequencies outside the lock range (not shown), we observe, as expected~\cite{Razavi2004}, \emph{injection pulling} of the optomechanical oscillator. This term comes from the fact that, in this range of detuning, the resonance peak appears to be pulled towards the injection signal~\cite{Razavi2004}.  

The locking phenomena discussed above are qualitatively the same as those observed for injection locking of electrical tank circuits~\cite{Razavi2004} as well as a previous demonstration with an optomechanical system~\cite{Hossein-Zadeh2008}. However, when these parameters are varied to obtain a larger locking range,  we find that the end-of-range dynamics are strikingly different.

Decreasing $P_\mathrm{in}$ to 10~mW with the same drive strength ($V_\mathrm{AC}=0.5~$V) to obtain a larger locking range~\cite{amitai2017pra}, we repeat the search for the ends of the lock range. With these settings, the amplitude of mechanical oscillations due to the radiation-pressure force and due the inertial drive are of the same order. We find that the end of the lock range is no longer marked by the distinct quasi-lock spectrum. Fig.~\ref{fig:Edgeoflock}(c) shows the spectra near the lower end of the lock range, where the drive frequency is decreased from the top to bottom traces. The end of the lock range is now instead marked by an increase in the power at the natural resonance frequency, rather than the quasi-lock spectrum. Outside the lock range here, injection pulling is not observed. We refer to this apparently distinct regime of partial locking as the `continuous-suppression' regime, as the peak at the mechanical resonance frequency appears to be suppressed as the injection signal moves into the locking range. The transition from the quasi-lock regime to the continuous-suppression regime is consistent with experimental parameters that cause the driven oscillation amplitude to approach or even exceed the amplitude due to optical pumping. The cross-over is not entirely well defined, however, as some settings result in one end of the lock range being marked by quasi-locking while the other being marked by continuous suppression of the natural peak.

In order to understand these two seemingly distinct locking regimes, we modelled the system using classical equations of motion for the coupled optical mode (characterized by its light amplitude $\alpha$) and mechanical mode (characterized by its position $x$) which can be derived from the Hamiltonian given earlier:

\begin{align}
\dot{\alpha} &= - \frac{\kappa}{2} \alpha + i(\Delta + Gx) \alpha + A_\mathrm{L} \\
m_\mathrm{eff}&[\ddot{x} + \Gamma \dot{x} + \omega_\mathrm{m}^2 x] = \hbar G \lvert \alpha \rvert^2 + F_\text{d}(t)
\label{eq:eom}
\end{align}

The mechanical mode, with decay rate $\Gamma/2\pi=15$~kHz, is subject to a radiation-pressure force and the electrical drive $F_\mathrm{d}(t)=(\delta C / \delta x)V_\mathrm{DC}V_\mathrm{AC}\cos(\omega_\mathrm{d} t)$. We omit in our simulations thermo-optic effects, which affect the refractive index of the material~\cite{Gil-Santos2017}, as well as the thermal Langevin force~\cite{Aspelmeyer2014}. We numerically solve the coupled differential equations to simulate the dynamics of the system in the locking regimes explored experimentally in Fig.~\ref{fig:Edgeoflock}(a,c), and plot the results in Fig.~\ref{fig:Edgeoflock}(b,d). We find excellent qualitative agreement between simulation and experiment, highlighting that these simple classical equations of motion capture all the experimentally observed locking dynamics, including the cross-over between the quasi-lock and continuous-suppression regimes.

\subsection{Quantifying the Locking Range}

\begin{figure} [!t]
	\includegraphics[width=\linewidth]{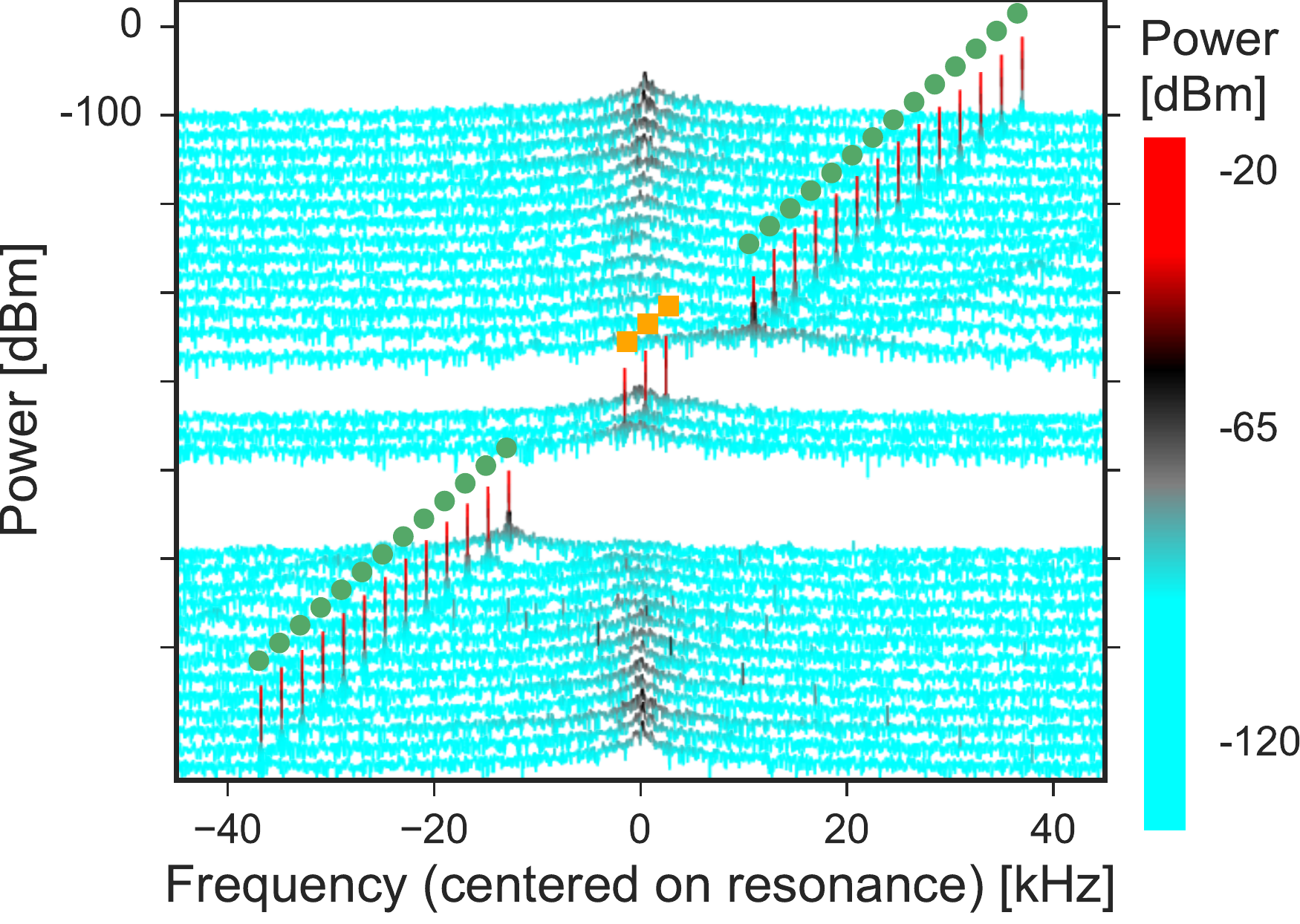}
	\caption{\label{fig:Spanningrange} Demonstration of the large lock range achieved in the experiment: 71~kHz as compared to the 30~mHz linewidth with $P_\mathrm{in}=11$~mW, with traces offset for clarity. The traces marked by green dots correspond to locking with $V_\mathrm{AC}=5$~V where $\omega_\mathrm{d}$ is varied. This large drive cannot be used close to $\omega_\mathrm{m}$ due to thermo-optic effects as explained in the Supplementary Materials. Nevertheless, decreasing $V_\mathrm{AC}$ allows the oscillations to be locked over the entire lock range, as demonstrated by the traces marked by orange squares where $V_\mathrm{AC}$ is reduced to 0.5~V.}
\end{figure}

In this section, we characterize the dependence of the lock range on optical power and drive strength and compare our data to the predictions of Amitai \textit{et al.}~\cite{amitai2017pra}. We first determine the largest lock range achievable with our device, by maximizing the output of the RF source to $V_\mathrm{AC}=5$~V and lowering the optical power to $P_\mathrm{in}=11$~mW. Fig.~\ref{fig:Spanningrange} plots traces of the power spectrum for varying drive frequencies that span the lock range for these settings, marked by green dots. The lock range achieved is 71~kHz, over $2\times 10^6$ times the linewidth of the unlocked regenerative oscillations. While this range can easily be increased by increasing $V_\mathrm{AC}$ or $V_\mathrm{DC}$, this value already represents a tuning percentage of $\sim 1\%$. For applications that require arrays of optomechanical systems regeneratively oscillating at the same frequency, it is important to compare this tuning percentage to the expected variations in natural mechanical resonance frequencies. For the case of a circular optomechanical resonator, the resonance frequency of a radial breathing mode is inversely proportional to the radius, $\omega_\mathrm{m} \propto 1/R$. Given that a 100~$\mu$m disk can be fabricated to well within 1~$\mu$m precision with standard lithographic techniques, the tuning range demonstrated here is already sufficient to overcome the $<1\%$ variations expected in $\omega_\mathrm{m}$.

Referring back to the locked traces in Fig.~\ref{fig:Spanningrange}, we find that setting the injection frequency of the strong drive ($V_\mathrm{AC}=5$~V) close to the mechanical resonance frequency causes the optical cavity to be shifted out of resonance such that regenerative oscillations cease. This is shown by the absence of traces marked by green dots between $-10$~kHz and 10~kHz. We suspect that this is due to thermal effects in the optomechanical system, as explained in the Supplementary Materials. Here, we emphasize that this effect does not preclude locking over the entire range, as $V_\mathrm{AC}$ can be reduced to lock near the natural mechanical resonance frequency. This is demonstrated by the power spectra marked by orange squares in Fig.~\ref{fig:Spanningrange}(c), where $V_\mathrm{AC}$ is reduced to 0.5~V to successfully lock near the centre of the range. 

Amitai \textit{et al.}~\cite{amitai2017pra} show theoretically that the locking range $2\omega_\mathrm{r}$ is directly proportional to the drive strength and inversely proportional to the amplitude of mechanical oscillations $r_o$ (units of m). Adapting their expression to our parameters, we have

\begin{equation}
\label{eq:Bruder}
\omega_\mathrm{r} = \frac{x_\mathrm{zp}^2}{\hbar}\frac{\delta C}{\delta x}\frac{V_\mathrm{AC}V_\mathrm{DC}}{r_o}.
\end{equation}

\noindent The amplitude of regenerative oscillations $r_o$ is a function of the optomechanical parameters of the system as well as $P_\mathrm{in}$, which can be controlled in the experiment. We first determine the lock range as a function of $P_\mathrm{in}$. As explained previously, two regimes of end-of-range dynamics are observed in the experiment. In one, $\omega_\mathrm{r}$ can be unambiguously quantified owing to the appearance of the quasi-lock spectrum. While some hysteresis was observed depending on the direction of the frequency sweep~\cite{Zalalutdinov2003}, its effect on quantifying $\omega_\mathrm{r}$ was minor and thus neglected. In the continuous-suppression regime, however, the end of the lock range is marked by a relatively gradual re-emergence of the peak at the natural frequency $\omega_\mathrm{m}$. For consistency, we choose to define $\omega_\mathrm{r}$ in this regime to be the frequency offset at which the peak at $\omega_\mathrm{m}$ reaches 30~dB below the peak at $\omega_\mathrm{d}$. We note that this choice of threshold does not have a significant effect on the results as the peak at $\omega_\mathrm{m}$ rises from completely suppressed (below the noise floor) back to its original amplitude when $\omega_\mathrm{d}$ is varied over a frequency offset range that is $\sim5\%$ of the locking range.

\begin{figure} [!t]
	\includegraphics[width=\linewidth]{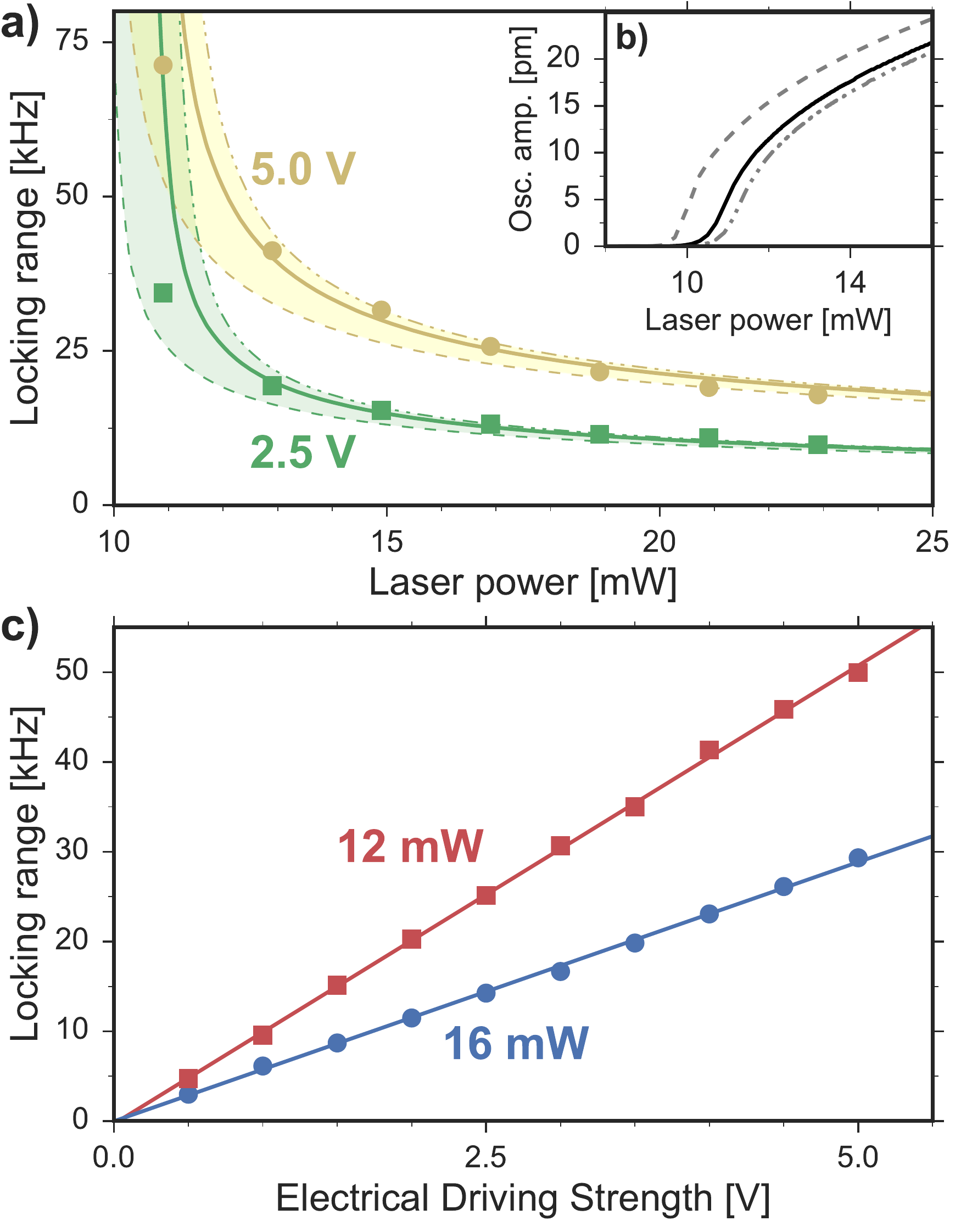}%
	\caption{\label{fig:Dependence} \textbf{(a)} Locking range as a function $P_\mathrm{in}$ for $V_\mathrm{AC}=2.5$~V (green squares) and 5~V (yellow circles). The fits to the measured locking ranges use the results of calculated mechanical oscillation amplitude $r_o$ as a function of $P_\mathrm{in}$, plotted in \textbf{(b)} for $\Delta/2\pi=42$~MHz (dash-dot), 44~MHz (solid) and 50~MHz (dashed). Using Eq.~\ref{eq:Bruder}, the resulting calculated locking ranges are correspondingly plotted in (a), with the best fit provided by $r_o(P_\mathrm{in},\Delta)$ with $\Delta/2\pi = 44$~MHz, and upper and lower bounds with 42~MHz and 50~MHz, respectively. The data lies within the shaded regions defined by the bounds, indicating good agreement between experiment and theory. \textbf{(c)} Locking range as a function of $V_\mathrm{AC}$ for $P_\mathrm{in}=12$~mW (red squares) and 16~mW (blue circles), following the linear dependence described by Eq.~\ref{eq:Bruder}.
	}
\end{figure}

Fig.~\ref{fig:Dependence}(a) plots the measured locking range $2\omega_\mathrm{r}$ as a function of $P_\mathrm{in}$ for $V_\mathrm{AC}=2.5$~V (green squares) and $5$~V (yellow circles). While this data includes measurements in both the quasi-lock and continuous-suppression regimes, the data points nevertheless follow clear trends. In order to compare these results to locking ranges predicted by Eq.~\ref{eq:Bruder}, we require the regenerative oscillation amplitude $r_o$, which is not precisely determined in the experiment. We thus use the equations of motion (Eq.~\ref{eq:eom}) to calculate $r_o$ in the absence of an inertial drive. Fig.~\ref{fig:Dependence}(b) plots $r_o$ as a function of optical power $P_\mathrm{in}$ for detunings $\Delta/2\pi=42$, 44 and 50~MHz, illustrating the sensitivity of $r_o$ to both these parameters. We find that, when inserted into Eq.~\ref{eq:Bruder}, $r_o(P_\mathrm{in},\Delta)$ for $\Delta/2\pi = 44$~MHz (solid line) produces the best fit to the measured locking ranges. The fit is plotted as the solid curve in Fig.~\ref{fig:Dependence}(a), demonstrating good agreement between theory and experiment. Note that a free scaling parameter $\beta = 0.65$ is added to Eq.~\ref{eq:Bruder}, to account for uncertainties in $\delta C/\delta x$ and $x_\mathrm{zp}$. The fit could further be improved by taking into account the modification of effective optical detuning with $P_\mathrm{in}$. Increasing $P_\mathrm{in}$ modifies the effective cavity resonance due to the steady-state radiation-pressure-induced cavity enlargement. This dependency is naturally corrected for in the experiment by always tuning the laser frequency to maximize the regenerative oscillation signal. However, for simplicity, the calculations for Fig.~\ref{fig:Dependence}(b) were performed with a fixed laser frequency. Adding this correction, which we estimate to result in a few MHz shift in detuning over the range of laser powers used in our experiments, would reduce the steepness of the calculated curves at lower laser power, resulting in a closer match to the experimental data. As confirmation of this mechanism, we include the calculated locking ranges for $\Delta/2\pi=42$~MHz and 50~MHz to Fig.~\ref{fig:Dependence}(a), which enclose all the experimental data. 

In contrast, the measurement of locking range as a function of $V_\mathrm{AC}$ does not involve changes in the optical detuning and can thus be more straightforwardly compared to the theory. Fig.~\ref{fig:Dependence}(c) plots this for $P_\mathrm{in}=12$~mW (red squares) and 16~mW (blue circles), confirming the linear dependence predicted by Eq.~\ref{eq:Bruder}.

\section{Conclusion}

We have reported the first observation of locking of the radiation-pressure driven regenerative oscillations of an optomechanical system to a direct inertial drive. We demonstrate a suppression of the phase noise of the regenerative oscillations of over 55~dBc/Hz at 2~Hz and a locking range of 71~kHz, more than 2~million~times the 30~mHz oscillation linewidth. This tuning range is sufficient to overcome variations in natural mechanical resonance frequencies due to limits in fabrication precision. Applications that require distributed optomechanical systems or arrays of oscillators within a single chip may benefit from injection locking. While phase-locked loops can also be used to stabilize the mechanical resonance frequency, they require feedback circuitry which may prove impractical for large arrays of devices. The feed-forward nature of injection locking allows stabilization to be achieved without feedback circuitry. Moreover, the electrical drive presented here can be implemented for arrays more efficiently than optical injection which requires either additional lasers or optical power modulators. This is especially true in the case where multiple oscillation frequencies are desired. 

At a system level, the direct inertial drive that we implement is distinct to that used in prior explorations of injection locking in optomechanical systems as well as in other physical platforms. To our knowledge, the only previously known difference between the two drive forms is the absence of harmonic locking effects for the direct drive~\cite{Zalalutdinov2003}. We present previously unreported locking dynamics as described by the continuous-suppression regime, enabled by the ability to increase the inertial injection signal to approach and even exceed that of the radiation-pressure force. Further research is required to explain the underlying mechanisms responsible for the cross-over between the two regimes with distinct locking dynamics. Recently, Toth \textit{et al.}~\cite{toth2017pla} also demonstrated an injection signal that bypasses the non-linearity in a radiation-pressure electromechanical system. This experiment explored the reverse regime to that studied here, with the mechanical oscillator driving regenerative oscillations of a microwave field, allowing injection locking of the microwave resonance.

Beyond applications in sensing, the electrical drive technique introduced in this work may enable the possibility to engineer optomechanical gain competition. For instance, we observed that for some devices, non-ideal circularity split the radially symmetric mechanical mode into two quasi-degenerate modes separated by $<1$~kHz, as visible in Fig.~\ref{fig:Edgeoflock}(c). Upon application of the drive to the resonance frequency of either mode, the system could be made to regeneratively oscillate on the targeted mode. Indeed, much like for the optical modes of a laser, different mechanical modes compete for gain in an optomechanical resonator~\cite{Kemiktarak2014}. We provide a detailed numerical analysis of this phenomenon in the Supplementary Materials and confirm that the electrical drive can reorient the oscillator along a new stable trajectory. The switch of optomechanical gain from one mode to the other persists even once the drive has been turned off, potentially serving as a form of `non-volatile' optomechanical memory~\cite{bagheri_dynamic_2011} or enabling the exploration of normally inaccessible stable dynamical attractors of the system \cite{marquardt_dynamical_2006, krause_nonlinear_2015}.





\vspace{3mm}
\textbf{Acknowledgments} This research was primarily funded by the Australian Research Council and Lockheed Martin Corporation through the Australian Research Council Linkage Grant LP140100595. Support was also provided by a Lockheed Martin Corporation seed grant and the Australian Research Council Centre of Research Excellence for Engineered Quantum Systems (CE110001013). W.P.B. and R.K. acknowledge fellowships from the Australian Research Council (FT140100650) and the University of Queensland (UQFEL1719237), respectively. This work was performed in part at the Queensland node of the Australian National Fabrication Facility, a company established under the National Collaborative Research Infrastructure Strategy to provide nano and microfabrication facilities for Australia's researchers.
\newline \\

\textbf{Additional Information} Supplementary information accompanies the paper. Correspondence and requests for materials should be addressed to C.B. (c.baker3@uq.edu.au) or W.B. (wbowen@physics.uq.edu.au).
\clearpage

\onecolumngrid
\begin{center}
	\textbf{SUPPLEMENTARY INFORMATION for \\
		``Injection locking of an electro-optomechanical system", by C. Bekker \emph{et.al.}\\
		\vspace{3mm}}
\end{center}
\twocolumngrid

\renewcommand\thesubsection{S\arabic{subsection}}
\setcounter{page}{1}
\renewcommand*{\thepage}{S\arabic{page}}
\setcounter{equation}{0}
\renewcommand{\theequation}{S\arabic{equation}}
\setcounter{figure}{0}
\renewcommand{\thefigure}{S\arabic{figure}}
\renewcommand*{\citenumfont}[1]{S#1}
\renewcommand*{\bibnumfmt}[1]{[S#1]}

\renewcommand\thesection{}

These supplements contain a detailed investigation into the effects of the inertial drive on optomechanical gain competition (section \ref{modecompetitionsection}). We also provide additional details on the phase offset between the mechanical oscillations and injection signal (\ref{sectionphase}), the observed instability at large drive amplitudes (\ref{section instability}) and the experimental setup (\ref{setup1} \& \ref{setup2}).

\subsection{Optomechanical mode competition}
\label{modecompetitionsection}

Here we employ the numerical model described in the main text to investigate  mechanical mode competition in an optomechanical resonator. Indeed, much like for the optical modes of a laser, different mechanical modes  may compete for gain in an optomechanical resonator \cite{kemiktarak_mode_2014}. We show here how the electrical drive technique introduced introduced in this work and \cite{baker_high_2016} may be used to permanently reorient the resonator's phase-space trajectory in order to, for instance, favor optomechanical gain for one mechanical mode versus another.

In order to illustrate this phenomenon, we consider for our simulation an optomechanical resonator with two distinct mechanical modes, chosen here to be at 6 MHz (mode 1) and 10 MHz (mode 2). We rewrite Eqs. 4 \& 5 of the main text, to account for the presence of the second mechanical mode:
  
\begin{align}
\dot{\alpha} &= - \frac{\kappa}{2} \alpha + i(\Delta + G_1 x_1 + G_2 x_2)\, \alpha + A_L \label{eq:2modes1}\\
m_\mathrm{eff,1}\,&[\ddot{x_1} + \Gamma_1 \dot{x_1} + \omega_{m,1}^2 x_1] = \hbar G_1 \lvert \alpha \rvert^2 + F_\text{d}(t) \label{eq:2modes2}\\
m_\mathrm{eff,2}\,&[\ddot{x_2} + \Gamma_2 \dot{x_2} + \omega_{m,2}^2 x_2] = \hbar G_2 \lvert \alpha \rvert^2 + F_\text{d}(t),
\label{eq:2modes3}
\end{align}
where the subscripts 1 and 2 respectively refer to mechanical modes 1 and 2. Both mechanical modes may have a distinct effective mass $m_{\mathrm{eff}}$, damping rate $\Gamma$, mechanical frequency $\omega_m$ and optomechanical coupling rate $G$, but both couple to the same intracavity field $\alpha$.

\begin{figure}[!h]
	\centering
	\includegraphics[width=\linewidth]{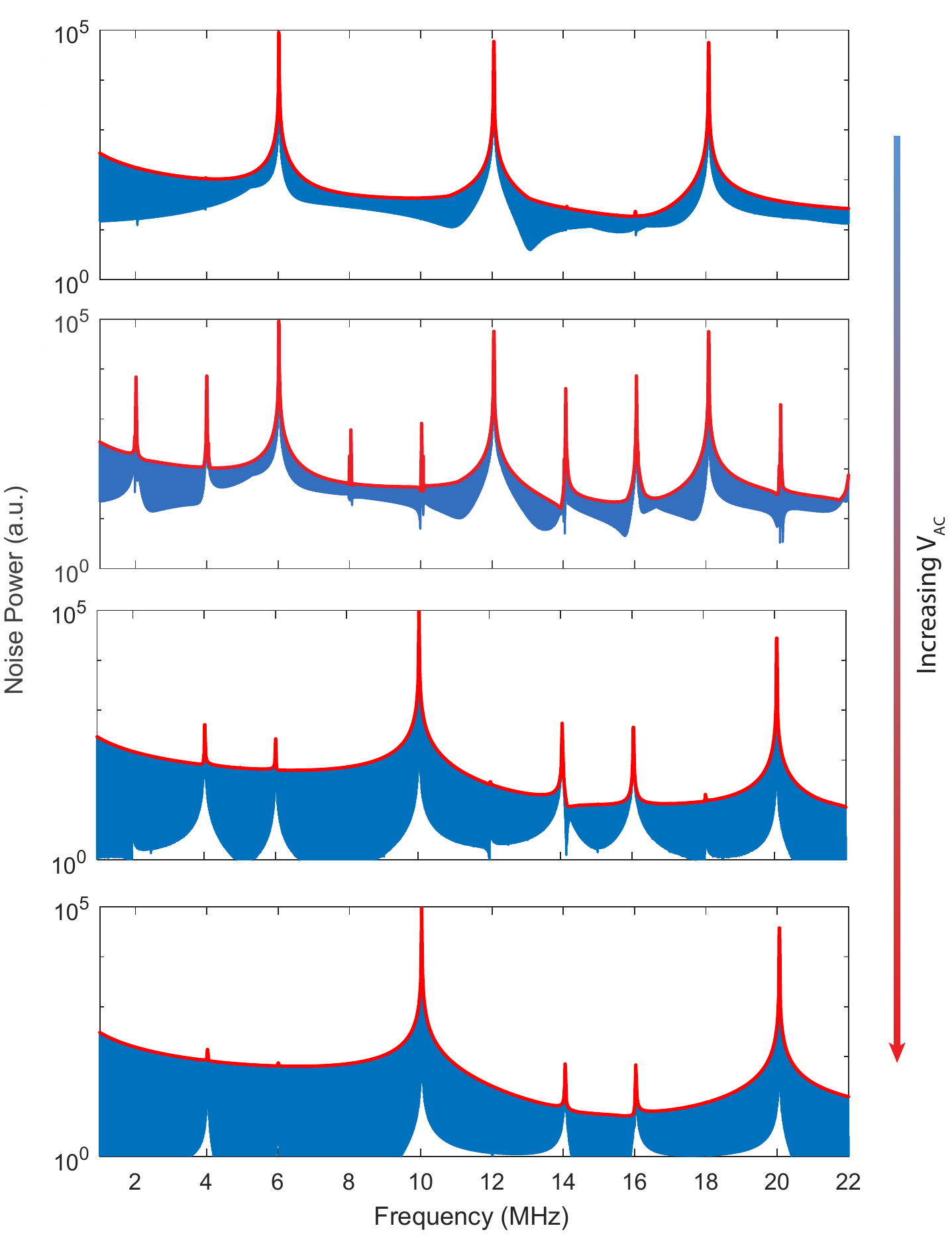}%
	\caption{\label{fig:1supChris} Noise power spectra obtained through the FFT of the simulated normalized cavity output intensity (blue line). Red line (moving maximum) serves as a guide to the eye. Simulation parameters are as follows: $Q_{\mathrm{opt}}=2 \times 10^6$ (loaded); $\omega_{m,1}/2\pi=6$ MHz; $\omega_{m,2}/2\pi=10$ MHz; $\Gamma_1/2\pi=6$ kHz; $\Gamma_2/2\pi=10$ kHz (mechanical Q of both modes $=1\times10^3$); $G_{1}/2\pi=G_{2}/2\pi=1.9\times 10^{18} $ Hz/m; $m_{\mathrm{eff},1}=m_{\mathrm{eff},2}= 5\times 10^{-11}$ kg; laser power = 5 mW; $\Delta/2\pi=$ 35 MHz; $dC/dx=2\times 10^{-9}$ N/V$^2$; $V_{\mathrm{DC}}=100$ V. From top to bottom panels, $V_{\mathrm{AC}}=0; 0.04; 0.05; 0.1$ V. Total simulation time 600 $\mu$s. In this simple model, we consider the effect of the drive F(d) to be identical on both mechanical modes, see Eqs. (\ref{eq:2modes2}) \& (\ref{eq:2modes3}). Peaks appearing at e.g. 4 and 16 MHz correspond to sum- and difference frequency terms between modes 1 and 2. Simulation code available from the authors upon request.}
\end{figure}

Figure \ref{fig:1supChris} demonstrates the effect of the external capacitive drive $F(d)$ 
on the optomechanical gain of each mechanical mode.
The intracavity photon number $\lvert \alpha \rvert^2 $ and cavity detuning $\Delta$ in the simulation are set such that each mechanical mode, taken individually, is above its regenerative oscillation threshold \cite{aspelmeyer_cavity_2014}. However, in the absence of drive ($V_{\mathrm{AC}}=F_d=0$), mode 1 'wins' the gain competition and is the only mode to reach regenerative oscillation, i.e. `phonon-lasing' (Fig. \ref{fig:1supChris}, top panel). This is visible by the large peaks at $\omega_{m,1}/2\pi$; $2 \omega_{m,1}/2\pi$ and $3 \omega_{m,1}/2\pi$ corresponding respectively to the fundamental mechanical frequency and first two higher order harmonics which appear due to the nonlinearity of the lorentzian optical resonance for large displacements  \cite{aspelmeyer_cavity_2014}. Note also the absence of any peak at $\omega_{m,2}/2\pi$. Next, the drive force $F_d(t)$ is set to the frequency of mode 2 (10 MHz), and its amplitude is progressively increased ($V_{\mathrm{AC}}=0.04; 0.05; 0.1$ V). The simulation results are plotted Fig. \ref{fig:1supChris}, second, third and fourth panels. For the highest drive voltage, the initial situation is completely reversed, with the noise peak corresponding to mode 2 now more than 30 dB above mode 1.

\begin{figure*} 
	\includegraphics[width=\textwidth]{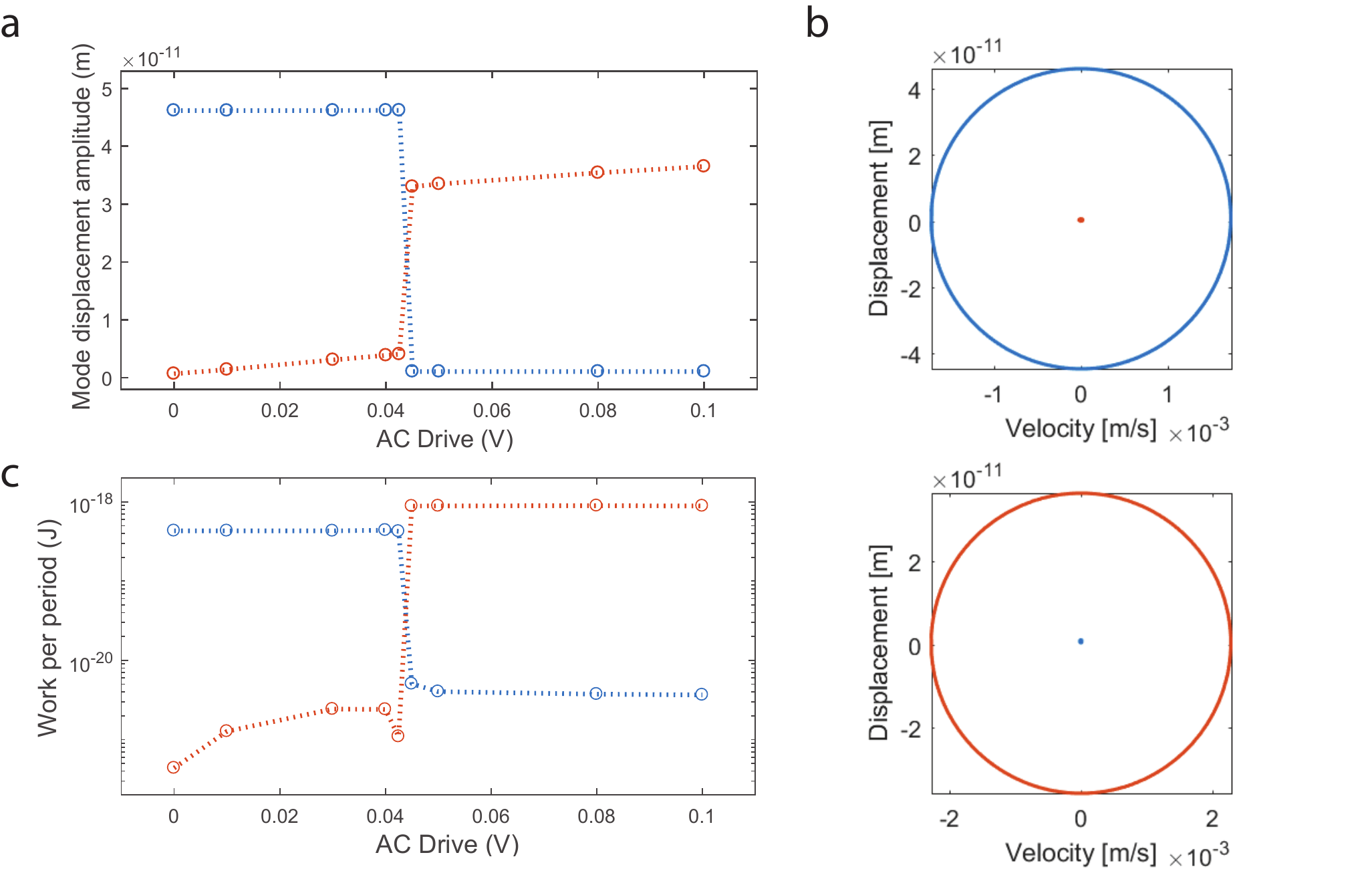}%
	\caption{\label{fig:2supChris}(a) Simulated steady-state mechanical displacement amplitude of mode 1 (blue) and mode 2 (orange), as a function of AC drive voltage. (b) Phase-space trajectories of mode 1 (blue) and mode 2 (orange), for no capacitive drive ($V_{\mathrm{AC}}=0$ V, top) and $V_{\mathrm{AC}}=0.1$ V (bottom). (c) Work $W_{\mathrm{rad}}$ performed by the intra-cavity optical field on mode 1 (blue) and mode 2 (orange) during one oscillation period. This corresponds to a net power transferred from the light field to the lasing mechanical mode on the order of a few picoWatts.}
\end{figure*}

The transition between the two extreme cases is more quantitatively presented in Fig. \ref{fig:2supChris}(a), which presents the displacement amplitude of each mechanical mode in the steady-state, as a function of drive voltage. The transition from mode 1 to mode 2 having the largest displacement amplitude occurs abruptly, for $0.0425<V_{\mathrm{AC}}<0.045$ V. Similarly, Fig. \ref{fig:2supChris}(b) plots the phase-space trajectories of mode 1 (blue) and mode 2 (orange), for no capacitive drive ($V_{\mathrm{AC}}=0$ V, top) and $V_{\mathrm{AC}}=0.1$ V (bottom).

At this stage, we seek to verify that the switch from `phonon-lasing' on mode 1 to mode 2 is indeed due to a change in the optomechanical gain experienced by both modes, and not to spurious effects due to the drive itself. In order to do so, we calculate the work done by the radiation pressure force on each mode during a mechanical oscillation. The radiation pressure force  acting on mechanical mode $i$ is given by $F_{\mathrm{rad},i}=\hbar G_i \,\lvert \alpha \rvert^2$, \cite{aspelmeyer_cavity_2014} such that the work performed by the intra-cavity optical field  on mode $i$ during one period is given by:

\begin{equation}
W_{\mathrm{rad},i}=\int_0^{T_i} \hbar G_i \,\lvert \alpha \rvert^2 \times \dot{x}_i
\end{equation}

This work is calculated from our time-domain simulation, once the system has reached the steady-state. Results for modes 1 and 2 are plotted in Fig. \ref{fig:2supChris}(c), as a function of $V_{\mathrm{AC}}$. We observe that the shift from `phonon-lasing' on mode 1 to mode 2 is indeed accompanied by a more than two orders of magnitude change in the radiation-pressure work received by both modes (decrease for mode 1; increase for mode 2). This confirms that the change in dominant mechanical resonance is indeed due to  gain competition between the two modes.
We further verify that the work $W_{\mathrm{d},i}$ done by the drive force
\begin{equation}
W_{\mathrm{d},i}=\int_0^{T_i} F\left(d \right) \times \dot{x}_i,
\end{equation}
remains always at least one order of magnitude below the radiation pressure work $W_{\mathrm{rad},i}$.

Finally, we investigate what happens when the electrical drive is turned off after switching from mode 1 to mode 2. Does the system decay back to oscillating mainly on mode 1, or does it exhibit sufficient hysteresis to keep oscillating on mode 2? By replacing the drive term $F_d(t)$ in Eqs. (\ref{eq:2modes2}) and (\ref{eq:2modes3}) by $F_d(t)\, H(t_{\mathrm{stop}} - t)$, where H is the Heaviside function, we turn off the electrical drive at a time $t_{\mathrm{stop}}$ once the system has been switched from mode 1 to mode 2. Even with the drive turned off, the  system maintains regenerative oscillation on mode 2, indicating that the capacitive drive has effectively reoriented the resonator along a new stable oscillation trajectory, favoring optomechanical gain for a different mechanical mode. 

The addition of an external driving means to a cavity optomechanical system may therefore enable dynamically adjusting the mechanical lasing mode, serve as a form of `non-volatile' mechanical memory \cite{bagheri_dynamic_2011} (provided the optical input is not removed), and enable the exploration of normally inaccessible stable dynamical attractors of the system \cite{marquardt_dynamical_2006, krause_nonlinear_2015}.

\subsection{Phase Offset of Locking and Edge-of-Lock Effects}
\label{sectionphase}

When a mechanical resonator is injection-locked, the phase of its motion maintains a constant offset with respect to the phase of the injection signal. The value of this phase offset depends on the detuning of the injection signal from the natural mechanical resonance frequency, as mentioned in the main text. We performed numerical simulations to investigate this dependence in each of the \emph{quasi-lock} and \emph{continuous} edge-of-lock regimes.

Our simulations consisted in solving the coupled differential equations that govern the time evolution of the optical and mechanical degrees of freedom (Eqs. 4 \& 5 of the main text). These were solved using an ODE solver (MATLAB) to determine the dynamics of the optomechanical system over a time period on the order of $\sim$1000 $\mu$s, with a few million time steps --corresponding to an average nanosecond-range time increment between successive points. Note that these simulations, even though simplified (they do not include thermo-optic effects or a random thermal force for instance), are nevertheless able to accurately reproduce our experimental observations (see Fig. 5 of the main text). This holds for both the quasi-lock and continuous regimes. The simulation code is available from the authors upon request.

\begin{figure} [h]
	\includegraphics[width = \linewidth]{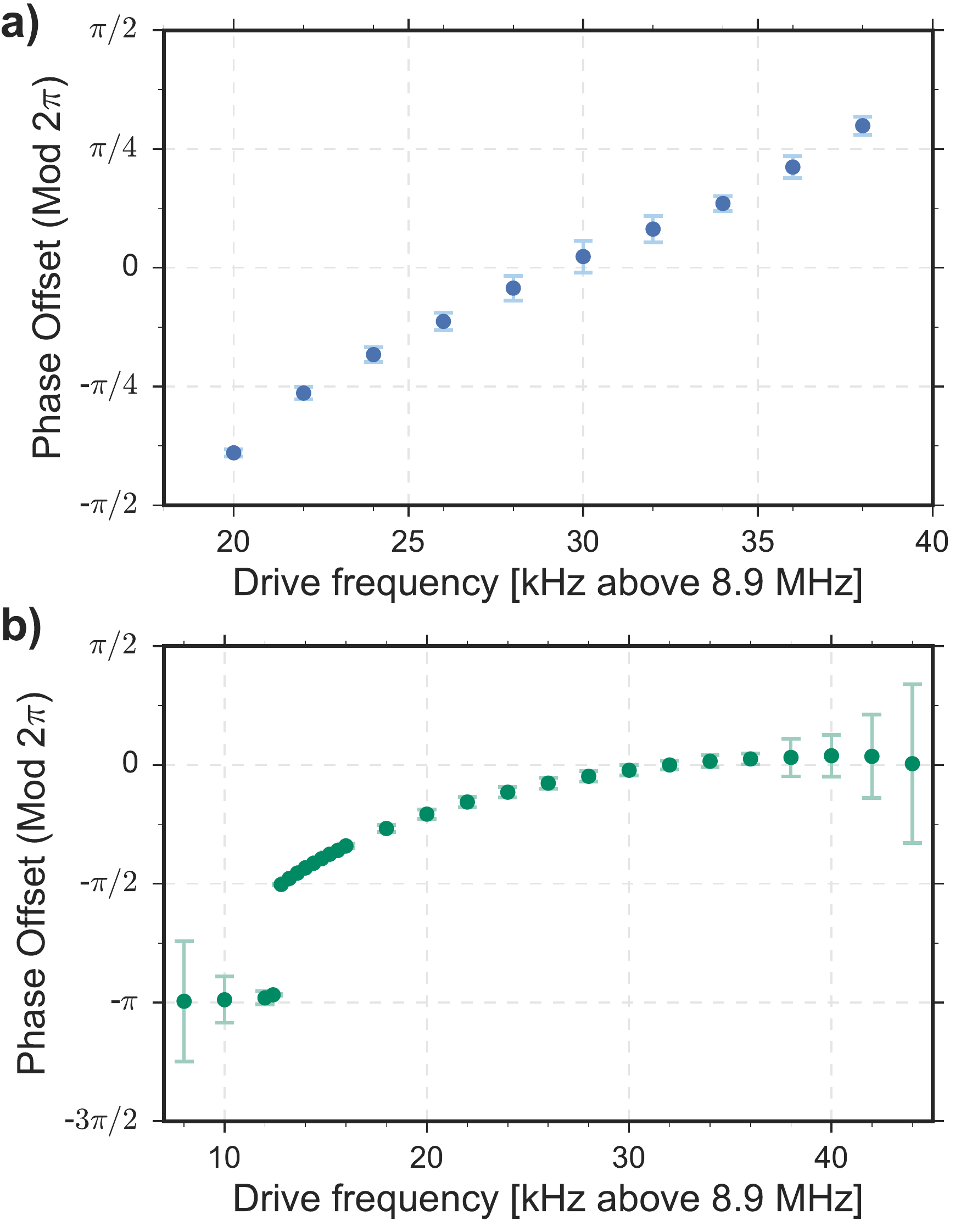}
	\caption{\label{fig:Offset1} Phase offset between the injection signal and the locked oscillations as a function of the drive frequency. The phase offset is measured between the electrical drive and the resonator position $x$, once the system has reached its steady-state in the simulation. These were numerically modelled in the \textbf{a)} quasilock and \textbf{b)} continuous regimes. Note how the behaviour of the phase offset is qualitatively different between the two regimes.}
\end{figure}

Our numerical modelling reveals that the dependency of the phase offset on the frequency of the injection tone is qualitatively different between the two regimes, as shown in Fig. \ref{fig:Offset1}. Figure \ref{fig:Offset1}(a) plots the phase offset between electrical drive and locked mechanical oscillator, as function of drive frequency, for simulation parameters corresponding to the quasi-lock regime. The locking here spans approximately from 8.920 MHz to 8.938 MHz, corresponding to a locking range of 18 kHz. 
The phase offset is in very good agreement to that expected from the theory of an injection-locked tank circuit \cite{Razavi2004}. The phase offset spans the expected $[-\pi/2;\pi/2]$ range, with zero phase offset in the center of the lock range, and an offset approaching $\pm\pi/2$ at the edges of the lock range \cite{Razavi2004}.

\begin{figure} [h]
	\includegraphics[width = \linewidth]{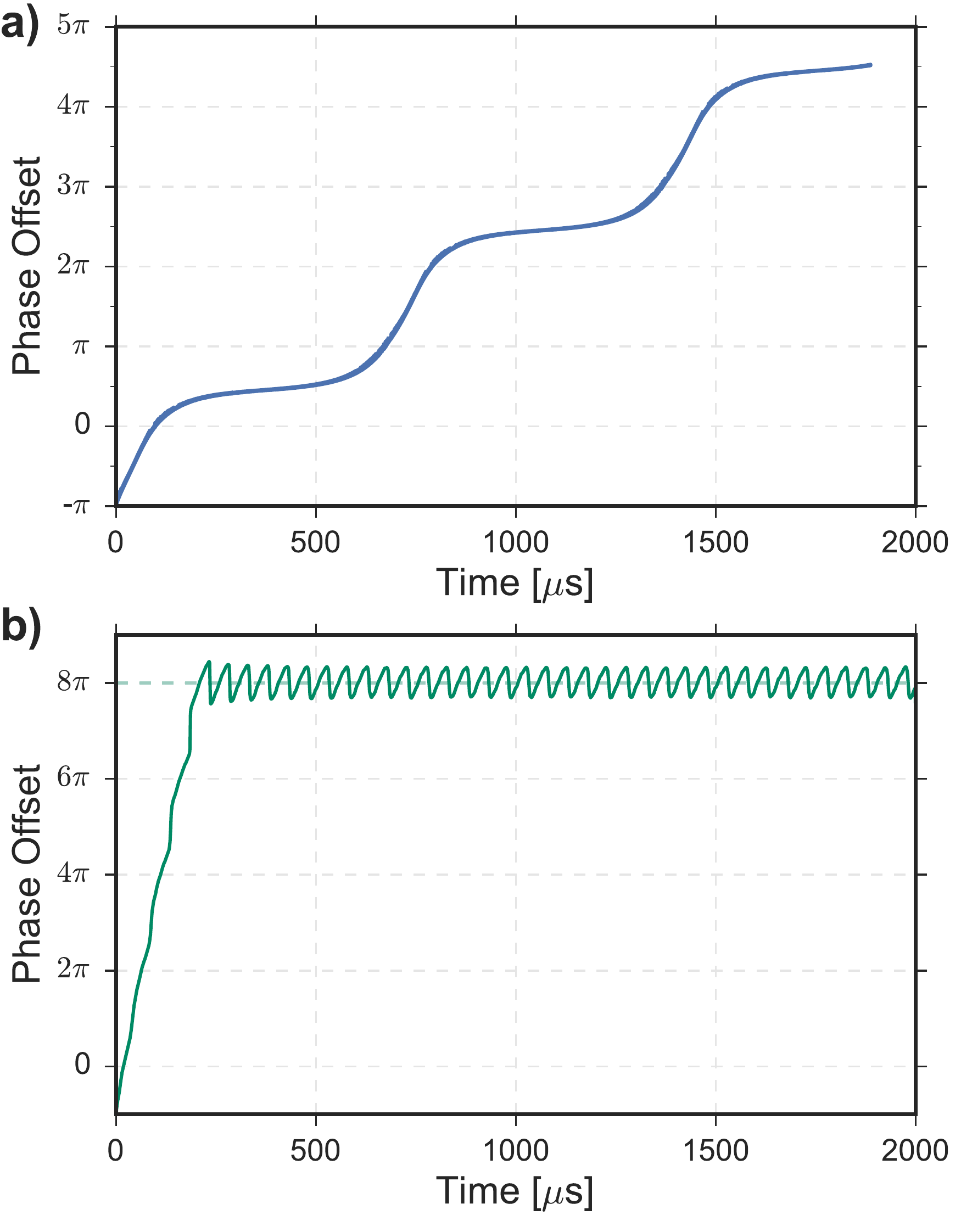}
	\caption{\label{fig:Time1} Phase offset as a function of time between the mechanical oscillator position $x$ and the injection signal. 
	a) Simulated phase offset in the quasilock regime for a drive frequency just outside the lock range (8.940 MHz).  The quasi-locking regime corresponds to the `phase-slip' regime described by Razavi \cite{Razavi2004}, where the oscillator is being held at a phase offset of $\pi/2$ modulo $2\pi$ for periods of time, before incurring a $2\pi$ phase slip.  b) Phase offset in the continuous regime for a drive frequency just on the edge of the lock range (8.944 MHz - rightmost point in Fig. \ref{fig:Offset1}(b)) . The green dotted line corresponds to a phase offset of $0$ modulo $2\pi$, shown in Fig. \ref{fig:Offset1}(b).}
\end{figure}

In striking contrast to the quasilock case, in the continuous regime the phase offset is bound within $[-\pi; 0]$, with a non-symmetric relationship of the phase-offset with respect to the center of the locking range.
Another unique feature of this regime is the apparently `instantaneous' jump of the phase offset from $-\pi$ to $-\pi/2$ close to the lower end of the locking range (8.913 MHz).  Near the edge of the locking range, the phase offset oscillates in time around a steady-state value, as shown in Fig. \ref{fig:Time1}(b). This oscillation is responsible for the increased uncertainty in defining the value of the phase offset, as evidenced by the growing error bars.

We now look at the behaviour of the phase offset in the time domain, for a drive frequency just on the edge of the lock range. This analysis reveals a qualitatively different mechanism through which the locking is lost in both quasi-lock and continuous regimes. Figure \ref{fig:Time1} plots the phase offset  between the mechanical oscillator position $x$ and the injection signal as a function of time. In Fig. \ref{fig:Time1}(a) the phase offset is shown for a drive frequency just outside the lock range, corresponding to a drive at 8.940 MHz in Fig. \ref{fig:Offset1}(a). Since this drive is outside of the locking range, the phase offset is no longer constant over time.  Nevertheless, the phase offset is maintained at $\pi/2$ modulo $2\pi$ for sustained periods of time (horizontal sections of the curve), before incurring a $2\pi$ phase slip, and locking again. The mechanical oscillator therefore alternates between periods of time where it is locked to the drive, and periods where it is not. This regime is therefore named the \emph{quasi-lock} or \emph{phase-slip} regime \cite{Razavi2004}. As the drive frequency is further moved away from the edge of the locking range, the fraction of the time the oscillator is locked versus slipping keeps diminishing. This is the mechanism whereby lock is lost in the quasilock regime.

In contrast, Fig.~\ref{fig:Time1}(b) plots the phase offset in the continuous regime for a drive frequency just on the edge of the lock range (corresponding to the rightmost point in Fig. \ref{fig:Offset1}(b)). The simulation shows an initially unlocked oscillator which locks after four $2\pi$ phase slips. After locking, no further phase-slips occur. Nevertheless, the phase offset oscillates about the value of $8\pi$ (corresponding to 0 mod $2\pi$ in Fig.~\ref{fig:Offset1}(b)). The further the drive frequency is moved away from the center of the lock range, the larger these phase oscillations become (as illustrated Fig.~\ref{fig:Offset1}(b)), until lock is lost. This analysis therefore underlines the qualitative differences between the quasi-lock and continuous regimes, both in terms of the phase offset inside the locking range, as well as the mechanism whereby the oscillator falls out of lock. Further research must be done to investigate other differences between these regimes, and uncover the precise mechanisms behind such qualitatively different behaviour.

\subsection{Instability of Large Drives}
\label{section instability}

As shown in the main text, we attempt to tune the regenerative oscillation frequency across a 71~kHz locking range using a $V_\mathrm{AC}=5$~V drive. However, we find that when the frequency of this strong drive is set close to the mechanical resonance frequency, the optical cavity is shifted out of resonance and regenerative oscillations cease. On the other hand, when the frequency of the drive is set away from resonance, regenerative oscillations remain stable and successfully lock to the drive tone. We suspect that this occurs due to a combination of large amplitude mechanical oscillations when driven near resonance along with thermal effects in the silica microtoroid~\cite{mcrae2009optx}. Silica has an appreciable thermo-optic coefficient, which describes the change in its refractive index with temperature. In effect, the optical resonance frequency shifts as the optical power inside the cavity is increased, resulting in optical bi-stability~\cite{almeida2004optical}. These shifts are routinely observed in the experiment through frequency scans of the optical mode, which produce a characteristic triangular shape~\cite{Rosenberg2009}. When driven near resonance to high amplitudes, the mechanical oscillations can sufficiently shift the optical resonance away from the laser frequency ($Gx>\kappa$) such that the optical power inside the cavity drops significantly. This results in a run-away effect where the silica cools as the resonance frequency continues to shift further away from the laser frequency until it reaches its equilibrium temperature, where the laser frequency is completely detuned from resonance. 

\subsection{Electric Probe Setup}
\label{setup1}
Electrical contact with our device is established through firmly pressing ultrasharp tungsten probes with tip radii of 1 $\mu$m onto the contact pads at the center of the resonator. We observe that, over time, the probes cause wear to the pads with each application, particularly when a DC voltage is applied across the electrodes. Tungsten is a very porous material, and oxide tends to build up on and within the probe. This accumulation of oxide can create a high contact resistance between the probe and the pad, with the potential for an applied voltage to cause arcing \cite{Accuprobe}, which we suspect is the cause of the observed wear. This problem may be avoided by using beryllium copper (BeCu) probe tips. These are softer than tungsten tips and are effectively polished when pressed against a surface. As the BeCu probe tip slides, the scrubbing action leaves the tip surface clean, maintaining its low contact resistance with the gold pads~\cite{Accuprobe}.

\subsection{Fibre Taper Stabilisation}
\label{setup2}

All measurements shown in the main text are performed with the tapered section of the fibre brought into contact with the top of the silica microtoroid. This is done out of necessity, as we are unable to achieve stable coupling of the optical cavity to the fibre while simultaneously applying a DC voltage to the electrodes and keeping the fibre positioned in proximity to the device. Invariably, the fibre drifts and is pulled towards the device in timescales on the order of one minute, likely due to polarization of the fibre and a resulting attractive force. With the voltage switched off, however, we are able to take measurements to extract the mechanical linewidth without the fibre touching. This comparison reveals that bringing the fibre into contact with the device only reduces the mechanical quality factor by a factor of $\sim3$. Fortunately, the presence of the fibre on top of the microtoroid does not severely damp the radial motion of the mechanical resonator. With the fibre in contact, stable regenerative oscillations are maintained for hours. 

For future devices, there are a number of ways in which stable optical coupling can be achieved without bringing the fibre into contact with the microtoroid. For instance, separate support structures such as nanoforks or support pads can be fabricated to stabilize the fibre~\cite{Rosenberg2009, mcauslan_microphotonic_2016}. Alternatively, the tapered fibre can be removed completely and integrated waveguides for optical coupling can be used instead~\cite{Tallur2011,baker_critical_2011}.


\end{document}